\documentstyle[prl,aps,epsf,multicol,tabularx]{revtex}
\tighten
\begin{document}
\title{Charge Fluctuations in the Edge States of N-S hybrid
Nano-Structures.}
\author{Andrew M. Martin, Thomas Gramespacher and Markus B\"uttiker}
\address{D\'epartement de Physique Th\'eorique, Universit\'e de Gen\`eve,
CH-1211 Gen\`eve 4, Switzerland.}
\date{\today}
\maketitle
\begin{abstract}
In this work we show how to calculate the equilibrium and non-equilibrium
charge fluctuations in a gated normal mesoscopic conductor which is attached
to one normal lead and one superconducting lead. We then consider an example
where the structure is placed in a high magnetic field, such that the
transport is dominated by edge states. We calculate the equilibrium and
non-equilibrium charge fluctuations in the gate, for a single edge state,
comparing our results to those for the same system, but with two normal
leads. We then consider the specific example of a quantum point contact and
calculate the charge fluctuations in the gate for more than one edge state.

\end{abstract}
\begin{multicols}{2}
\narrowtext
\section{Introduction}
Recently a methodology for calculating charge fluctuations
in gated hybrid normal-superconducting systems has been formulated
\cite{ammtgmb}. In
this Letter we implement this methodology for one particularly relevant
and enlightening example.
We study the system depicted in Fig. 1, i.e. a gated N-S system in
a high magnetic field, where the transport in the normal region is governed by
edge states. We focus on
calculating the charge fluctuations in the edge states in the region
$\Omega$. This work coincides with recent experimental and theoretical research
into the properties of hybrid normal-superconducting structures in high
magnetic fields \cite{Takayanagi,Ma,Ishikawa,Schon}.

In this Letter we  first calculate
the unscreened charge fluctuations for a general N-S structure (Sec. 2).
Then we proceed to incorporate screening into the problem (Sec. 3),
enabling us to calculate the spectra of charge fluctuations in a gated N-S
structure. We show that these spectra, at equilibrium, are determined by
two quantities: the charge relaxation resistance, $R_q$, and the electrochemical
capacitance $C_{\mu}$. In the presence of transport, we find that $R_q$ is
replaced by $R_V$, called here the Schottky resistance, which reflects the
shot noise of the structure. Then
we focus on one particular example (Sec. 4) an N-S structure in a
high magnetic field, see Fig. 1. In particular we are interested in the
charge fluctuations in the edge states in the region $\Omega$ of Fig. 1
which can be measured by observing the current fluctuations at the gate
$S_{II}(\omega,V)=\omega^2 S_{QQ}(\omega,V)$. We
consider a general case of a single edge state and an arbitary scatterer and
more than one edge state in the case where the scatterer is a quantum point
contact.
\section{Calculating the unscreened charge fluctuations}

Consider a conductor where we have one normal lead and one superconducting
lead, we can write an effective 
\begin{figure}
\narrowtext
\epsfxsize=7cm
\centerline{\epsffile{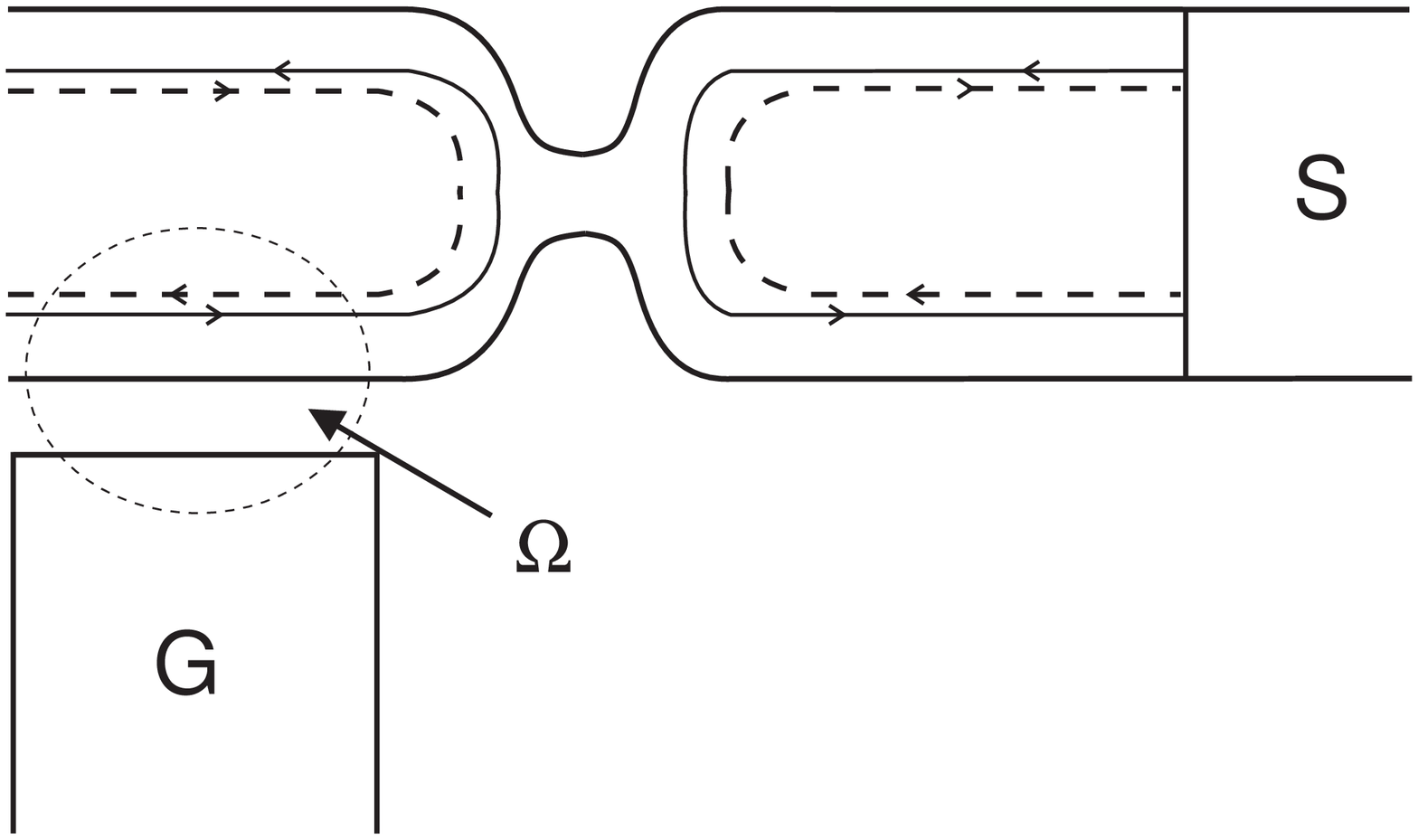}}
\vspace{0.2cm}
\caption{Quantum point contact, in the presence of a high magnetic
field, attached to one normal lead and a superconducting lead,
capacitively coupled to a macroscopic gate. The solid (dashed)
lines depict the {\it motion} of the electron (hole) edge states}
\label{Fig1a}
\end{figure}
\noindent $2 \times 2$ scattering matrix for this
system,
\begin{equation}
    \hat{S} = \left( \begin{array}{ll}

        s_{pp} & s_{hp} \\

        s_{ph} & s_{hh}

        \end{array} \right)
        \label{seff},
\end{equation}
where the $s_{\mu \lambda}$ are elements of the scattering matrix describing
the process of an electron ($\lambda=p$) or hole ($\lambda=h$) entering from
the normal lead and an electron ($\mu=p$) or hole ($\mu=h$) returning to the
normal lead. Making use of this $\hat{S}$-matrix
\cite{ammtgmb,Thomas99} we can write down the local particle density of
states elements
at position ${\bf r}$ as
 \begin{eqnarray}
{\cal N}_{\alpha \beta}^{\eta}(\lambda, {\bf r})
&=& \frac{-1} {4\pi i}  \left[
s_{\lambda \alpha}^{\dagger}(E,\underline{U}({\bf
r})) \frac{
\partial s_{\lambda \beta}(E,\underline{U}({\bf r}))}
{q^{\eta} \partial U^{\eta}({\bf r})} \right. \nonumber \\  & - &
\left.  s_{\lambda \alpha}(E,\underline{U}({\bf
r})) \frac{
\partial s_{\lambda \beta}^{\dagger}(E,\underline{U}({\bf r}))}
{q^{\eta} \partial U^{\eta}({\bf r})} \right] \label{pdos}
\end{eqnarray}
where $\alpha$, $\beta$, $\lambda$ and $\eta$ denote the
electron/hole degrees of freedom ($p/h$) and $q^{p}=e=-q^{h}$. The
functional derivatives are taken at the equilibrium electrostatic
potential $U^{p} = U^{h} = U_{eq}$. To give an example, ${\cal
N}_{ph}^{h}(h, {\bf r})$ is the hole density associated with
an electron and a hole current amplitude incident from the normal lead
and reflected back into the normal lead as an outgoing hole amplitude.
With the help of this basic expression we can now find both the average
density of states as well as the fluctuations.
The total bare charge fluctuations in the conductor are given by
\begin{equation}
{\cal N}_{\alpha \beta}= -{\cal N}_{\alpha
\beta}^h+ {\cal N}_{\alpha \beta}^p
\label{fluct}
\end{equation}
where
\begin{equation}
{\cal N}_{\alpha \beta}^{\eta}= \int_{\Omega} d^3
{\bf r}\sum_{\lambda} {\cal N_{\alpha
\beta}^{\eta}(\lambda, \bf r}).
\label{phfluct}
\end{equation}
The hole density of
states of a region $\Omega$ of the conductor is
$N^h=N^h(p)+N^h(h)$ where
\begin{equation}
N^h(\alpha)= \sum_{\lambda}\int_{\Omega}d^3{\bf r}
{\rm Tr} [{\cal N}_{\alpha \alpha}^{h}(\lambda, {\bf
r})] \label{hdos}
\end{equation}
and the electron density of states is $N^p=N^p(p)+N^p(h)$ with
\begin{equation}
N^p(\alpha)= \sum_{\lambda} \int_{\Omega}d^3 {\bf r}
{\rm Tr} [{\cal N}_{\alpha \alpha}^p(\lambda, {\bf r})].
\label{ppdos}
\end{equation}
The trace is over open quantum channels,
$N^{\alpha}(\beta)$ is the injectivity of electrons (holes) ($\beta=p (h)$),
from the normal lead into the conductor, given a change in the particle (hole)
potential ($\alpha=p(h)$).

The fluctuations of the bare charge in a region $\Omega$ of interest
can be found from the charge operator
$e\hat{{\cal N}}$ given by
\begin{eqnarray}
 e \hat{{\cal N}} (\omega) &=&
\sum_{\alpha  \beta \atop \eta \lambda}
\int_{\Omega} d^3 {\bf r}
\int dE  \nonumber \\
&\times& [\hat{a}^{\dagger}_{\alpha} (E) q^{\eta}
{\cal N}^{\eta}_{\alpha \beta}
(\lambda, {\bf r}; E, E+\hbar \omega)
\hat{a}_{\beta} (E+\hbar \omega)]
\label{nfluct}
\end{eqnarray}
where the zero-frequency
limit of
${\cal N}^{\eta}_{\alpha \beta}
(\lambda, {\bf r}; E, E+\hbar \omega)$
is given by Eq. (\ref{pdos}).
In Eq. (\ref{nfluct}),
$\hat{a}^{\dagger}_{\alpha} (E)$ ($\hat{a}_{\alpha} (E)$) creates (annihilates)
an incoming electron/hole ($\alpha=p/h$) in the normal lead.
The true charge fluctuations must be obtained by taking into
account the Coulomb interaction and below we show how to obtain
the true charge fluctuations from the fluctuations of the bare
charges.

\section{Calculating the total charge fluctuations}

Consider an N-S system where we have a gate, Coulomb coupled, to
a region, $\Omega$, of the normal part of the system. Then, assuming the gate
has no dynamics of its own, the charge in the region coupled to the gate
is given by $\hat{Q}=C\hat{U}$ where $\hat{U}$ is
the operator for the internal potential of the conductor. Also we can
determine the charge $\hat{Q}$ as the sum of the bare charge fluctuations
$e\hat{{\cal N}}$ and the induced charges generated by the fluctuating
induced electrical potential. In the random phase approximation the induced
charges are proportional to the average frequency-dependent density of
states $N_{\Sigma}(\omega)$ times the fluctuating potential. Hence, the net
charge can be written as
$\hat{Q}=C\hat{U}=e\hat{{\cal N}}-e^2N_{\Sigma}(\omega)\hat{U}$, this is the
direct analogue of the result for a system with two normal leads
\cite{Pedersen}. Thus
\begin{equation}
\hat{U}=Ge\hat{{\cal N}}
\end{equation}
where
\begin{equation}
G(\omega)=\frac{1}{C+e^2N_{\Sigma}(\omega)}.
\end{equation}
and the total density of states is
\begin{equation}
N_{\Sigma}=\frac{1}{2}[N^p(p)-N^p(h)+N^h(h)-N^h(p)].
\end{equation}
We now wish to find the fluctuation spectra of the internal potential
\cite{Buttiker96}
\begin{equation}
2 \pi S_{UU}(\omega) \delta(\omega-\omega^{\prime})= (1/2)
\langle \hat{U}(\omega)\hat{U}(\omega^{\prime})+
\hat{U}(\omega^{\prime})\hat{U}(\omega) \rangle.
\end{equation}
Solving the above and making use of the fact that
$C^2 S_{UU}(\omega,V)= S_{QQ} (\omega,V)$ we find
\begin{eqnarray}
S_{QQ}(\omega,V)  &=& (1/2)
C_{\mu}^2 N_{\Sigma}^{-2}
\sum_{\alpha \beta}
\int
dE F_{\alpha \beta}(E,E+\hbar \omega)  \nonumber \\
& \times & {\rm Tr}
[{\cal N}_{\alpha \beta}(E, E+\hbar \omega)
{\cal N}_{\alpha \beta}^{\dagger}(E, E+\hbar
\omega)]
\label{eq:fluct}
\end{eqnarray}
where
\begin{eqnarray}
 {\cal N}_{\alpha \beta} (E, E^{\prime}) &=&
\sum_{\eta \lambda}  {\rm sgn} (q^{\eta}) \int_{\Omega} d^3 {\bf r}
{\cal N}^{\eta}_{\alpha \beta}
(\lambda, {\bf r}; E, E^{\prime}), \\
 C_{\mu} &=&\frac{Ce^2 N_{\Sigma}}{C+e^2N_{\Sigma}}
\end{eqnarray}
and
\begin{equation}
F_{\alpha \beta}(E, E^\prime) =
f_{\alpha}(E)[1-f_{\beta}(E^\prime)] + f_{\beta}(E^\prime)[1-f_{\alpha}(E)].
\end{equation}
In the above the sum is over all degrees of freedom $\alpha \beta$,
the Fermi functions
$f_{\alpha}(E)$ are defined such that
$f_{p}=f_0(E_p-\mu)$ and
$f_{h}=f_0(E_h+\mu)$ where $E_{\alpha}$ is the
energy of a particle (hole) ($\alpha= p(h)$) in the normal reservoir,
which is at a chemical potential $\mu$,
$f_0(E)$ is the Fermi function at the condensate chemical
potential of the superconducting lead ($\mu_0$).

Now we wish to evaluate Eq. (\ref{eq:fluct}) at equilibrium and zero
temperature, to leading order in $\hbar \omega$. We find
\begin{equation}
S_{QQ}(\omega)=2 C_{\mu}^2 R_q \hbar |\omega|
\end{equation}
with an equilibrium charge relaxation resistance
\begin{equation}
R_q=\frac{h}{2e^2}\frac{\sum_{\alpha \beta} {\rm Tr}(
{\cal N}_{\alpha \beta} {\cal N}_{\alpha \beta}^{\dagger})}{[N_{\Sigma}]^2}.
\end{equation}
Also we find at zero temperature to leading order in $e |V|$ that
\begin{equation}
S_{QQ}(V)=2C_{\mu}^2 R_V  e |V|
\end{equation}
with a Schottky resistance
\begin{equation}
R_V=\frac{h}{2e^2}\frac{{\rm Tr}
({\cal N}_{p h} {\cal N}_{p h}^{\dagger}+
{\cal N}_{h p} {\cal N}_{h p}^{\dagger})}{[N_{\Sigma}]^2}.
\end{equation}

\section{Example}
We now wish to consider the structure shown in figure 1. If
there is only one edge state then we can proceed to consider the
situation when the gate
only {\it sees} the edge state in the area $\Omega$. In this
case a particle (hole) transversing the length {\it seen} by the
gate acquires a phase $\phi_p $ ($\phi_h$), this phase is
determined by the potential in this region. This simplifies the s-matrix
given by Eq. (\ref{seff}) to
\begin{equation}
\hat{S} = \left( \begin{array}{ll}

        r_o \exp{i\phi_p} & r_a \exp{i(\phi_p+\phi_h}) \\

        r_a^{\star} & r_o^{\star} \exp{i\phi_h}

        \end{array} \right),
\end{equation}
where $r_or_o^{\star}$ is the probability of normal reflection
($R_o^{NS}$) from the N-S structure, $r_ar_a^{\star}$ is the probability
of Andreev reflection ($R_a^{NS}$) and
$R_a^{NS}+R_o^{NS}=1$.
 We can then write, for the edge state described by the above scattering
 matrix, $ds/q^{\lambda} dU^{\lambda}= (d s/ d \phi_{\lambda}) ( d
\phi_{\lambda}/q^{\lambda} dU^{\lambda})$ and that $( d
\phi_{\lambda}/q^{\lambda} dU^{\lambda})=-2\pi dN_{\lambda}/dE$,
where $dN_{\lambda}/dE$ is the particle (hole) ($\lambda=p(h)$)
density of states in the normal lead. After some algebra and
assuming $dN_{h}/dE=dN_{p}/dE=N$ we can compare the
charge relaxation resistance for the N-S structure
($R_q^{NS}$) and the same structure but with both leads being
normal ($R_q^{N}$). We find
$R_q^{NS}=h/(2e^2R_o^{NS})$ and
$R_q^{N}=h/(2e^2)$. Hence in equilibrium we find for the spectra of the
charge fluctuations that
\begin{equation}
S_{QQ}^{NS}(\omega)=\frac{he^2N^2R_o^{NS}C^2}{(C+e^2NR_o^{NS})^2} \hbar |\omega|
\label{sqqw}
\end{equation}
and
\begin{equation}
S_{QQ}^{N}(\omega)=\frac{he^2N^2 C^2}{(C+e^2N)^2} \hbar |\omega|.
\end{equation}
In the prescence of an applied voltage
we find the Schottky resistances $R_V^{NS}=R_a^{NS}h/(2e^2R_o^{NS})$
and $R_V^{N}=R_o^N(1-R_o^N)h/e^2$, where $R_o^N=1-T_o^N$ is the
probability of normal reflection when we have replaced the
superconducting lead with a normal lead. Hence
the non-equilibrium charge fluctuations are given by
\begin{equation}
S_{QQ}^{NS}(V)=\frac{he^2N^2R_o^{NS}R_a^{NS}C^2}{(C+e^2NR_o^{NS})^2} e |V|
\label{sqqv}
\end{equation}
and
\begin{equation}
S_{QQ}^{N}(V)=\frac{he^2N^2 T_o^{N}R_o^{N} C^2}{(C+e^2N)^2} e |V|.
\end{equation}
The results above are for one edge state but the striking difference between
the results for a N-S system as compared to a system where both leads are
normal  occurs when more than one edge state is present. To consider this
scenario we shall focus on the particular example of our scattering region
being a quantum point contact. For such a system only one quantum channel
opens at a time, i.e. for all other open channels the transmission through
the quantum point contact is 1, in the case of an ideal superconducting
interface this implies that, for these open channels $R_a^{NS}=1$. Hence,
considering an additional edge
state which has perfect transmission through the quantum point contact then
this state generates no extra noise, and no extra contribution to screening,
i.e. the total added charge is zero ($C_{\mu}$ for this state is zero), thus
Eqs. (\ref{sqqw}) and (\ref{sqqv}) remain valid, where $R_o^{NS}$ and
$R_a^{NS}$ are the reflection probabilities for the opening quantum channel.
This is in contrast to what happens to $S_{QQ}^N(\omega)$ and $S_{QQ}^N(V)$
where, again no noise is added, but an extra screening charge is \cite{mbam}
for each edge state added, thus reducing the charge fluctuations seen in the
gate, as more edge states are introduced.
\section{Conclusions}
In this work we have shown how to calculate the screened charge fluctuations
in a gated N-S structure, we then proceeded to focus on the example of how
charge fluctuations in the edge states of an N-S system differ from those of a normal
system. The most startling difference is that for the example of a quantum
point contact scatterer the screened charge fluctuations are unaffected by the
presence of extra edge states. This is in contrast to the normal case
\cite{mbam}. This can be understood in the context of the electrochemical capacitance
which for the N-S system is given by
\begin{equation}
C_{\mu}^{NS}=\frac{e^2NR_o^{NS}C}{C+e^2NR_o^{NS}}.
\label{eq:cmuns}
\end{equation}
For perfect Andreev reflection $C_{\mu}^{NS}=0$ in contrast to the normal
system where the electrochemical capacitance is given by
\begin{equation}
C_{\mu}^{N}=\frac{e^2NC}{C+e^2N}
\end{equation}
which for finite $C$ and $N$ implies $C_{\mu}^{N} \ne 0$.
Comparing the two above equations we see that the introduction of Andreev
reflection reduces the electrochemical capacitance.
It has been shown elsewhere that fluctuations in the charge
in normal structures, as characterized by $R_q$ and $R_V$, determines the
dephasing rate \cite{mbam}. We expect that
such a relationship is also true for normal superconducting hybrid structures.
Finally induced currents have been observed in gates which are coupled to
normal conductors \cite{Chen}. Here we see, from
Eq. (\ref{eq:cmuns}), that
charging is reduced in the presence of Andreev reflection and in the
extreme case of perfect Andreev reflection the charging is zero.


This work was supported by the Swiss National Science Foundation
and by the TMR network Dynamics of Nanostructures.


\end{multicols}
\end{document}